\documentstyle[11pt]{article}

\newcommand{\BE}{\begin{equation}}
\newcommand{\EE}{\end{equation}}
\newcommand{\BA}{\begin{eqnarray}}
\newcommand{\EA}{\end{eqnarray}}
\newcommand{\half}{{\scriptstyle{\frac{1}{2}}}}
\begin{document}
\begin{titlepage}
\begin{flushright}
\hspace*{8.5cm} {\small DE-FG05-92ER40717-14}
\end{flushright}

\vspace*{14mm}
\begin{center}
               {\LARGE\bf ``Triviality'' Made Easy: \\
\vspace*{4mm}
                    the real $(\lambda\Phi^4)_4$ story}

\vspace*{18mm}
{\Large  M. Consoli}
\vspace*{5mm}\\
{\large
Istituto Nazionale di Fisica Nucleare, Sezione di Catania \\
Corso Italia 57, 95129 Catania, Italy}
\vspace*{5mm}\\
and
\vspace*{5mm}\\
{\Large P. M. Stevenson}
\vspace*{5mm}\\
{\large T.~W.~Bonner Laboratory, Physics Department \\
Rice University, Houston, TX 77251, USA}
\vspace{16mm}\\
{\bf Abstract:}
\end{center}

\par
The real meaning of ``triviality'' of $(\lambda\Phi^4)_4$ theory
is outlined.  Assuming ``triviality'' leads to an effective
potential that is just the classical potential plus the zero-point
energy of the free-field fluctuations.  This $V_{{\rm eff}}$ gives
spontaneous symmetry breaking.  Its proper renormalization has the
consequence that all scattering amplitudes vanish, self-consistently
validating the original assumption.  Nevertheless, the theory is
physically distinguishable from a free field theory; it has a
symmetry-restoring phase transition at a finite critical temperature.

\end{titlepage}

\setcounter{page}{1}

{\bf 1.}  The strong evidence that $\lambda \Phi^4$ theory is
``trivial'' in 4 dimensions \cite{triv,latt} seemingly conflicts
with the textbook description of the Standard Model, in which
$W$ and $Z$ masses arise from spontaneous symmetry breaking (SSB)
in the $\lambda \Phi^4$ scalar sector.  Current thinking holds
that the theory can only be saved by a finite ultraviolet cutoff,
thereby abandoning the one grand principle underlying the Standard
Model --- renormalizability.  In our view, ``triviality'' is true,
but its meaning and its consequences have not been properly understood.

    Our earlier papers \cite{cs} discuss the arguments in detail,
but here our exposition is as terse as possible so that the overall
picture can be seen whole.  The key point is this:  The effective potential
of a ``trivial'' theory is not necessarily a trivial quadratic function.
The effective potential is the classical potential plus quantum effects,
and in a ``trivial'' theory the only quantum effect is the zero-point
energy of the free-field vacuum fluctuations.

\vspace*{2mm}
{\bf 2.}  Consider the Euclidean action of classically-scale-invariant
$\lambda \Phi^4$ theory:
\BE
S[\Phi] = \int \!{\rm d}^4 x \left(
\frac{1}{2} \partial_{\mu} \Phi_B \partial_{\mu} \Phi_B
+ \frac{\lambda_B}{4!} \Phi_B^4 \right),
\EE
and substitute
\BE
\Phi_B(x) = \phi_B + h(x),
\EE
where $\phi_B$ is a constant.  (To make the decomposition unambiguous we
impose $\int \! {\rm d}^4 x \; h(x) = 0$ using a Lagrange multiplier $\eta$.)
Upon expanding one obtains $S[\Phi] = S_0 + S_1 + S_2 + S_{\rm int}$ where
\begin{eqnarray}
S_0 & = &  \frac{\lambda_B}{4!} \phi_B^4 \! \int \! {\rm d}^4 x ,
\\
S_1 & = &  \left( \frac{\lambda_B}{6} \phi_B^3  - \eta \right)
\int \! {\rm d}^4 x \, h(x),
\\
S_2 & = &  \int \! {\rm d}^4 x \left(
\frac{1}{2} \partial_{\mu} h \partial_{\mu} h +
\frac{1}{2} ({\scriptstyle \frac{1}{2}} \lambda_B \phi_B^2) h(x)^2 \right),
\\
\label{sint}
S_{\rm int} & = &\int \! {\rm d}^4 x \, \frac{\lambda_B}{4!} \left(
4 \phi_B h(x)^3 + h(x)^4 \right).
\end{eqnarray}
Consider the approximation in which we ignore $S_{\rm int}$.  It is
then straightforward to compute the effective action by the standard
functional methods.  Briefly, the linear term $S_1$ effectively plays
no role; the $S_0$ term simply reproduces itself in the effective
action; and the $S_2$ term reproduces itself together with a zero-point
energy contribution from the functional determinant.  Thus, the (Euclidean)
effective action is:
\BE
\label{Gamma}
\Gamma = - \int \! {\rm d}^4 x \left[
\frac{1}{2} \partial_{\mu} h \partial_{\mu} h
+  \frac{1}{2} ({\scriptstyle \frac{1}{2}} \lambda_B \phi_B^2) \, h(x)^2
+  V_{{\rm eff}}(\phi_B) \right],
\EE
where
\BE
V_{{\rm eff}}(\phi_B) = \frac{\lambda_B}{4!}\phi_B^4 +
\frac{1}{2} \int \!\! \frac{d^4p}{(2\pi)^4} \,
\ln (p^2 + {\scriptstyle \frac{1}{2}} \lambda_B \phi_B^2 ).
\EE
This effective action describes a free $h(x)$ field with a
$\phi_B$-dependent mass-squared, $\frac{1}{2}\lambda_B \phi_B^2$.
The effective potential for $\phi_B$ is just the classical
potential plus the zero-point energy of the $h(x)$ field.

[More precisely, the exact effective potential is the `convex envelope'
of this $V_{{\rm eff}}$; Ritschel's version of our calculation
shows explicitly how this comes about \cite{rit}.  $V_{{\rm eff}}$
is the usual ``one-loop effective potential''.
However, $\Gamma$ is {\it not} the one-loop effective action.  Our
approximation is not based on loop counting; it is defined by the
statement ``ignore $S_{\rm int}$''.]

\vspace*{2mm}
{\bf 3.}
   After subtracting a constant and performing the mass
renormalization so that the second derivative of the effective
potential vanishes at the origin, one has \cite{CW}:
\BE
\label{vq1}
V_{{\rm eff}} = \frac{\lambda_B}{4!}\phi_B^4 +
\frac{\lambda_B^2 \phi_B^4}{256 \pi^2} \left( \ln
\frac{\frac{1}{2} \lambda_B \phi_B^2}{\Lambda^2}
- \frac{1}{2} \right),
\EE
where $\Lambda$ is an ultraviolet cutoff.  This function is just
a sum of $\phi_B^4 \ln \phi_B^2$ and $\phi_B^4$ terms.  It has
a pair of minima at $\phi_B = \pm v_B$ and may be re-written in
the form:
\BE
\label{vq2}
V_{{\rm eff}} = \frac{\lambda_B^2 \phi_B^4}{256 \pi^2}
\left( \ln \frac{\phi_B^2}{v_B^2} - \frac{1}{2} \right).
\EE
Comparing the equivalent forms (\ref{vq1}) and (\ref{vq2}) gives
$v_B$ in terms of $\Lambda$.  Hence, the mass-squared of the $h(x)$
fluctuation field, $\frac{1}{2}\lambda_B \phi_B^2$, when evaluated in
the SSB vacuum, is
\BE
\label{mlam}
m_h^2 = {\scriptstyle \frac{1}{2}} \lambda_B v_B^2 = \Lambda^2
\exp \left( - \frac{32 \pi^2}{3 \lambda_B} \right).
\EE
Demanding that this particle mass be finite requires an infinitesimal
$\lambda_B$:
\BE
\lambda_B = \frac{32 \pi^2}{3} \frac{1}{\ln (\Lambda^2/m_h^2)} \to 0_+.
\EE
[This implies a negative $\beta$ function:
$\Lambda \partial \lambda_B / \partial \Lambda = - b_0 \lambda_B^2$,
with $b_0 = 3/16 \pi^2$.]

\par  It follows that $v_B$ goes to $\infty$, but the {\it depth} of
the SSB vacuum, $\lambda_B^2 v_B^4/512 \pi^2 = m_h^4/128 \pi^2$, remains
finite.  Thus, $V_{{\rm eff}}(\phi_B)$ becomes infinitely flat.  However,
the effective potential can be made manifestly finite by re-scaling
the constant background field $\phi_B$.  One defines $\phi_R$ as
$Z_{\phi}^{-1/2}\phi_B$, with $Z_{\phi} \propto 1/\lambda_B \to \infty$,
so that the combination
$\xi \equiv {\scriptstyle \frac{1}{2}} \lambda_B Z_{\phi}$
remains finite.  The physical mass is then finitely proportional to
$v_R = Z_{\phi}^{-1/2}v_B$; i.e., $m_h^2 = \xi v^2$.  The requirement
that the second derivative of $V_{{\rm eff}}$ with respect to $\phi_R$
at $\phi_R=v_R$ should be $m_h^2$ fixes $\xi$ to be $8 \pi^2$.  Thus, one
obtains:
\BE
\label{vex}
 V_{{\rm eff}}~ =
{}~\pi^2 \phi^4_R \left( \ln {{\phi^2_R}\over{v_R^2}}
 - {{1}\over{2}} \right),
\EE
and
\BE
\label{mex}
m^2_h~ = ~8\pi^2 v_R^2.
\EE

     Although the constant field $\phi$ requires an infinite re-scaling,
the fluctuation field $h(x)$ is not renormalized: in the effective action
(\ref{Gamma}) the kinetic term for $h(x)$ is already properly normalized.
The different re-scaling of the zero-momentum mode $\phi$ and the
finite-momentum modes $h(x)$ is the only truly radical feature of our
analysis.  We return to this issue in Sect. 5.

\vspace*{2mm}
{\bf 4.}
      What about the interaction term $S_{\rm int}$ that we neglected?
It generates a 3-point vertex $\lambda_B \phi_B$ and a 4-point vertex
$\lambda_B$.  Since our renormalization requires these to be of
order $\sqrt{\epsilon}$ and $\epsilon$, respectively (where
$\epsilon \sim 1/\ln \Lambda$, or $\epsilon=4-d$ in dimensional
regularization), these interactions are of infinitesimal strength.
This is true to {\it all} orders because any diagram with $T$ three-point
vertices, $F$ four-point vertices, and $L$ loops is, at most, of order
$(\sqrt{\epsilon})^T (\epsilon)^F (1/\epsilon)^L = \epsilon^{T/2+F-L}$.
It is a topological identity that $T/2+F-L = n/2 - 1$, where $n$ is the
number of external legs.  Hence, the full 3-point function vanishes like
$\epsilon^{1/2}$; the full 4-point function vanishes like $\epsilon$, etc.
Thus, we {\it obtain} ``triviality'' as a direct consequence of the way
we were obliged to renormalize the effective potential.  Our initial
approximation of ignoring the interaction terms $S_{\rm int}$ is seen
to be self-consistently justified because, physically, $S_{\rm int}$
produces no interactions.  Thus, our starting point is not
actually an approximation but rather an {\it ansatz} that produces a
{\it solution} of the theory.

     The subtlety, though, is that $S_{\rm int}$, while too weak to
produce physical interactions, can seemingly give contributions to
the propagator and to the effective potential.  The above
$\epsilon$-counting argument applied to the $n=2$ case implies that
there are finite contributions to the propagator from arbitrarily
complicated diagrams.  Similarly, in the $n=0$ case there are
${\cal O}(1/\epsilon)$ and finite contributions to the vacuum diagrams,
and hence to $V_{{\rm eff}}$.  However, our claim is that all of these
contributions will be re-absorbed by the renormalization process;
the unmeasurable quantities $\lambda_B, Z_{\phi}, v_B$, etc., may
change, but the physical results (\ref{vex}, \ref{mex}) will not.
This ``exactness conjecture'' is supported by three arguments: (i)
Since the theory has no physical interactions it would be paradoxical
for the effective potential to have a form other than that produced
by the classical potential plus free-field fluctuations.  How, physically,
can there be non-trivial contributions to the effective potential due
to interactions when, physically, there are no interactions?
(ii) In the Gaussian approximation, which accounts for all the ``cactus''
(``superdaisy'') diagrams generated by $S_{\rm int}$, one finds exactly
the same physical results (\ref{vex}, \ref{mex}).  Things are different
at the bare level, but the physical results are nevertheless exactly
the same \cite{zeit}.
(iii) The effective potential computed on the lattice in the
appropriate region of bare parameters agrees very nicely
with the one-loop form \cite{agodi}.  This is in spite of the fact
that $\lambda_B \ln \Lambda$ is of order unity in this region, and so
naively the two-loop contribution would be expected to be as large as
the one-loop contribution.

     Furthermore, simple diagrammatic arguments can immediately establish
part of the ``exactness conjecture''.  By $\epsilon$ counting it follows
that finite contributions to the 2-point function come only from terms
that gain a $1/\epsilon$ from {\it every} loop.
Such terms cannot depend on the external momentum $p$, so the
additional contributions only affect the mass renormalization.
Similarly, to obtain a net $1/\epsilon$ contribution from a vacuum
diagram, one must gain a $1/\epsilon$ from {\it every} loop.  Such
terms obey naive dimensional analysis and are proportional to $\phi^4$.
The associated sub-leading finite contributions will involve
$\phi^4 \ln \phi^2$.  However, one cannot obtain any other functional
dependence on $\phi$; terms with two or more powers of $\ln \phi$
will be suppressed by one or more powers of $\epsilon$.  Thus the
effective potential, at any order, is a sum of $\phi^4$ and
$\phi^4 \ln \phi^2$ terms.  It can therefore always be parametrized
as $A \phi^4(\ln (\phi^2/v^2) - {\scriptstyle \frac{1}{2}})$.  All that
one cannot show by this simple argument is that, after renormalization,
the coefficient $A$ must be $\pi^2$.

\vspace*{2mm}
{\bf 5.}
       As we have seen, the interactions of the $h(x)$ field vanish
because $\lambda_B \to 0$, but the effective potential is non-trivial
because there one has $Z_{\phi} \to \infty$ to compensate for
$\lambda_B \to 0$.  Thus, it is crucial for our picture that the
$Z^{1/2}_{\phi}$ re-scaling of the constant background field $\phi_B$
is quite distinct from the $Z^{1/2}_h = 1$ re-scaling of the fluctuation
field $h(x)$.  The decomposition $\Phi_B(x)=\phi_B+h_B(x)$, which
separates the zero 4-momentum mode from the finite-momentum modes,
is a Lorentz invariant decomposition for a {\it scalar} field.  Hence,
we can see no valid objection to treating the re-scaling of $\phi$ and
$h(x)$ separately.  The separation of the zero mode is particularly
straightforward and natural in a finite-volume context \cite{rit}.
The situation is directly analogous to Bose-Einstein condensation
where the lowest state must be given special treatment because it,
and it alone, aquires a macroscopic occupation number.

\par  $V_{{\rm eff}}$ is the generator of the zero-momentum Green's
functions \cite{CW}:
\BE
 V_{{\rm eff}}(\phi_R) =
 V_{{\rm eff}}(v_R)
-\sum^{\infty}_{n=2} {{1}\over{n!}}
\Gamma^{(n)}_R(0,0,...;v_R)(\phi_R-v_R)^n .
\EE
The $\Gamma^{(n)}_R$'s at {\it zero} momentum, being derivatives of the
renormalized effective potential, are finite.  However, at finite momentum,
the $\Gamma^{(n)}_R$'s vanish for $n \ge 3$, corresponding to
`triviality'.  This just means that the $p^{\mu} \to 0$ limit is not smooth:
The zero mode has non-trivial interactions, but the finite-momentum modes
do not.  The 2-point function is a special case: at finite momentum it is
$\Gamma^{(2)}_R(p) = p^2 + m_h^2$, which is the (Euclidean) inverse
propagator of a free field of mass $m_h^2$.  It {\it does} have a smooth
limit at $p^{\mu}=0$, because we required
\BE
\label{concon}
\left. {{d^2 V_{{\rm eff}}(\phi_R)}\over{d\phi^2_R}} \right|_{\phi_R=v_R}
 = m^2_h.
\EE
Physically, the point is this: The $h(x)$ fluctuations (which in
some sense are infinitesimal on the scale of $\phi_R$ if they were finite
on the scale of $\phi_B$) are sensitive only to the quadratic dependence of
$V_{{\rm eff}}$ in the immediate neighbourhood of $v_R$.  This quadratic
dependence should mimic the potential for a free field of mass $m_h$
for self consistency.

\vspace*{2mm}
{\bf 6.} Although this solution to $\lambda \Phi^4$ theory is ``trivial''
(meaning, it has no observable particle interactions), it is not
entirely trivial --- it is physically distinguishable from a free
field theory.  One can see this by considering finite temperatures.
The free thermal fluctuations add to $V_{{\rm eff}}$ a term
\BE
\frac{1}{\beta} \int \frac{{\rm d}^3 p}{(2 \pi)^3}
\; \ln [ 1- \exp \{- \beta (p^2 + 8 \pi^2 \phi_R^2)^{1/2} \} ],
\EE
where $\beta=1/T$.  This term leads to a first-order symmetry-restoring
phase transition at a finite, not an infinite, temperature:
$T_c = 2.77 v_R$ (i.e., $T_c = 0.31 m_h$).  It is $v_R$, not $v_B$,
that sets the scale because the {\it depth} of the SSB vacuum
(invariant under $\phi$ re-scalings) was $\half \pi^2 v_R^4$.
Thus, the non-trivial self-interactions of the zero mode, responsible
for the non-trivial shape of $V_{{\rm eff}}$, do reveal themselves
in the finite-temperature behaviour of the theory.

\vspace*{2mm}
{\bf 7.}  We have discussed only the $N=1$ theory, but everything can
be generalized to the O($N$)-symmetric case \cite{cs}.  There will be
$N-1$ massless, non-interacting Goldstone fields.  Their zero-point energy
is only an infinite constant, so the shape of the effective potential
should be identical to the $N=1$ case.  This is our second ``exactness
conjecture'' \cite{cs}.  It is supported by lattice evidence
\cite{agodi2} and by a non-Gaussian variational calculation \cite{rit2}.

   We considered only the classically-scale-invariant (CSI) theory
here, but everything can be generalized to include a bare
$\half m_B^2 \Phi_B^2$ term in the Lagrangian \cite{cs}.
However, not only is the CSI theory simpler, it is the most attractive
possibility \cite{CW}.  The only mass scale in the Standard Model
would be $v_R$, arising from dimensional transmutation.  One would
have a definite prediction for the Higgs mass; $m_h^2 = 8 \pi^2 v_R^2$,
which implies $m_h = 2.2$ TeV.  (There are relatively small corrections
due to the gauge and Yukawa couplings.  These couplings would also
induce weak interactions of the Higgs.)

   It is usually believed that a Higgs above 800 GeV is either
impossible \cite{latt} or must have a huge width and be associated
with strongly interacting longitudinal gauge bosons.  These
beliefs stem from the false notion that $m_h^2$ is proportional to
``$\lambda_R v_R^2$''.  ``Triviality'' means that the ``renormalized
coupling'' $\lambda_R$ vanishes; it does not mean that $m_h$ must also
be zero in the continuum limit \cite{huang}.

\vspace*{2mm}
\begin{center}
{\bf Acknowledgements}
\end{center}

\vspace*{-1.5mm}
This work was supported in part by the U.S. Department of Energy under
Grant No. DE-FG05-92ER40717.

\end{document}